\documentclass[prb,twocolumn,floatfix,superscriptaddress]{revtex4}

\usepackage{graphicx}
\usepackage{dcolumn}
\usepackage{bm}
\usepackage{amsmath,amssymb}
\usepackage{color}
\usepackage[normalem]{ulem}
\usepackage{braket}
\usepackage{simpler-wick}
\def\ket#1{| #1 \rangle}
\def\bra#1{\langle #1 |}

\begin{document}
\title{Multilevel resonant tunneling in the presence of flux and charge noise}

\author{Anatoly Y.~Smirnov}
\affiliation{D-Wave Systems Inc., 3033 Beta Avenue, Burnaby BC
Canada V5G 4M9}
\author{Alexander Whiticar}
\affiliation{D-Wave Systems Inc., 3033 Beta Avenue, Burnaby BC
Canada V5G 4M9}
\author{Mohammad H.~Amin }
\affiliation{D-Wave Systems Inc., 3033 Beta Avenue, Burnaby BC
Canada V5G 4M9}
\affiliation{Department of Physics, Simon Fraser University, Burnaby, BC Canada V5A 1S6}

\newcommand{\be}{\begin{equation}}
\newcommand{\ee}{\end{equation}}
\newcommand{\ba}{\begin{eqnarray}}
\newcommand{\ea}{\end{eqnarray}}
\newcommand{\nn}{\nonumber \\}

\begin{abstract}

Macroscopic resonant tunneling (MRT) in flux qubits is an important experimental tool for extracting information about noise produced by a qubit's surroundings. Here we present a detailed derivation of the MRT signal in the rf-SQUID flux qubit allowing for effects of flux and charge fluctuations on the interwell and intrawell transitions in the system. Taking into consideration transitions between the ground state in the initial well and excited states in the target well enable us to characterize both flux and charge noise source affecting the operation of the flux qubit. 
The MRT peak is formed by the dominant noise source affecting specific transition, with flux noise determining the lineshape of the ground to ground tunneling, whereas charge noise reveals itself as additional broadening of the ground to excited peak.
\end{abstract}

\maketitle

\section{Introduction}

An rf-SQUID flux qubit \cite{Lukens2000, AFL2000, MWJNature2011} is made of a ring of a superconducting wire interrupted by a Josephson junction. For the sake of tunability the single Josephson junction can be replaced by the compound Josephson junction (CJJ) forming a small additional loop \cite{Makhlin99,Harris2010} as it is shown in Fig.~1a. Applying an external flux bias $\Phi^x_{\rm CJJ}$ to the CJJ loop allows the tuning of an effective Josephson energy of the qubit. The current in the main loop can flow clockwise or counterclockwise corresponding to two minima (left and right wells) of the qubit's potential energy $U$ depicted in Fig.~1b. 
We use $\ket{0}$ and $\ket{1}$ to
denote the lowest energy metastable states in the left
and in the right wells, i.e., ground states in each well
in the absence of tunneling. Transitions between these
states are facilitated by quantum fluctuations.
An external flux $\Phi^x$ applied to the main loop moves  the bottoms of the left and right wells up or down with respect to each other (here we measure $\Phi^x$ with respect to $\Phi_0/2$, hence $\Phi^x = 0$ represents the degeneracy point). 
 The rate of transition exhibits a macroscopic resonant tunneling (MRT) peak when energies of the left-well, $E_0$, and the right-well, $E_1$,  ground states are aligned: $E_0 \simeq E_1$ as demonstrated in Refs.~\cite{AminAverin2008, HarrisMRT2008,LantingMRT2011}. Fluctuations $\delta \Phi$ of the external flux break the alignment of the left and right wells resulting in broadening of the MRT signal. 
 The understanding of noise spectrums from low- and high-frequency flux noise together with charge noise is critical in designing qubits with longer coherence times. 
 \cite{Yan2016,Quintana2017, Bylander2011}

In the present paper we discuss theoretical aspects of multilevel macroscopic resonant tunneling in the rf-SQUID, which includes transitions between two arbitrary states in the opposite wells.
In particular, we assume that the system is initialized in the left-well ground state $\ket{0}$ and subsequently tunnels to the right well. Applying 
the external flux $\Phi^x$ to the main loop of the rf-SQUID allows us to tune the energy $E_n$ of the target state $\ket{n}$ in resonance with $E_0$. The right-well states $\ket{n}$ are numbered by odd digits: $n=1,3,5,\ldots.$, with the first MRT peak corresponding to  $\ket{0} \rightarrow \ket{1}$ transition. The second MRT peak is associated with $\ket{0} \rightarrow \ket{3}$ interwell transition complemented by $\ket{3}\rightarrow \ket{1}$ intrawell transition. 

We will show that charge noise makes a significant contribution to the intrawell relaxation leading to additional broadening of this peak. Charge noise plays an even more important role in the higher MRT peaks, where more channels of intrawell relaxation are available. 

Our goal is to derive an expression for the escape rate from the initial $\ket{0}$ as a function of the external flux $\Phi^x$. The lineshape of the peak is determined by combined effects of flux and charge noise. To reach this goal we generalize the hybrid-noise approach  \cite{ASMA2018} to include intrawell relaxation. 
Within this approach, coupling to low-frequency flux noise is treated nonperturbatively, whereas interactions of the rf-SQUID with high-frequency flux noise and with charge noise are considered as small perturbations. 
The present paper provides a  detailed theory for the model used in Ref.~\cite{AW2022} to describe experimental results of multilevel MRT. 

\section{rf-SQUID Hamiltonian}
\label{sectionSQUID} 

\begin{figure}
\begin{center}
\includegraphics[width=\columnwidth]{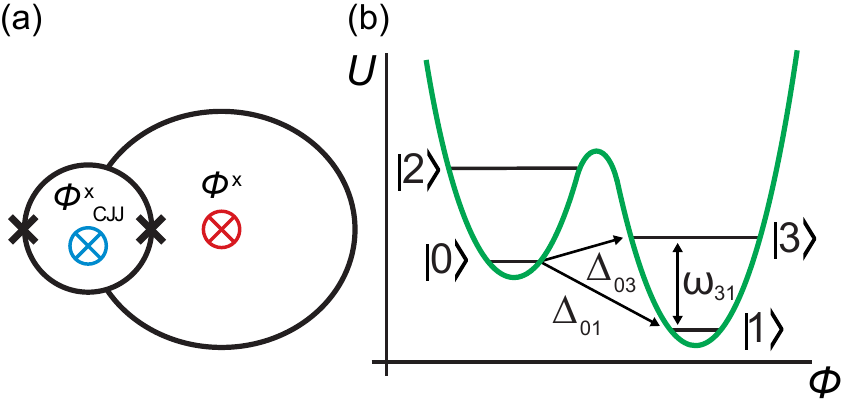}
\caption{\label{fig1} 
(a) 
An rf-SQUID flux qubit is comprised of the main loop
threaded by an external flux $\Phi^x$ and the CJJ loop threaded by an external flux $\Phi^x_{\rm CJJ}$. 
(b)  Potential energy $U$ of the  flux qubit as a function of the flux $\Phi$ induced in the main loop. We show four metastable energy levels, two in each well.}
\end{center}
\end{figure}

We consider a radio frequency superconducting quantum interference device (rf-SQUID) \cite{MWJNature2011, Harris2010} made of a  compound Josephson junction (CJJ). This device consists of 
two superconducting loops as shown in Fig.~\ref{fig1}(a). The CJJ loop has two Josephson junctions, with phase drops $\varphi_1$ and $\varphi_2. $ The junctions are characterized by critical currents $I_{C1}$ and $I_{C2}$ and capacitances $C_1$ and $C_2$. Hereafter we assume that
Josephson junctions forming the CJJ loop are symmetric, $I_{C1} = I_{C2}$, with the total critical current  $I_C = I_{C1}{+}I_{C2}$. 
Inductances of the main loop and the CJJ loop are denoted as $L$ and $L_{\rm cjj}$, respectively.
The Hamiltonian of  the system is written as
 \ba \label{Ha}
 H_{S}
 = \frac{q^2}{2C} + \frac{q_{\rm CJJ}^2 }{2C_{\rm CJJ}} + U(\varphi,\varphi_{\rm CJJ}), 
 \ea
where 
$C {=} C_1 {+} C_2$ and $1/C_{\rm CJJ} {=} 1/C_1 {+} 1/C_2$
are parallel and series combinations of the
junction capacitances, $q$ and $q_{\rm cjj}$ are the sum and difference of
the charges stored in these capacitors, respectively. The charges $q$ and $q_{\rm CJJ}$ are canonically conjugated to phases $\varphi {=} (\varphi_1 {+} \varphi_2)/2$ and $\varphi_{\rm CJJ} {=} \varphi_1 {-} \varphi_2$: $ [ \varphi, q ] {=} [ \varphi_{\rm CJJ}, q_{\rm CJJ} ] = 2ie $. 
Flux quantization relates the flux $\Phi$ threading the main loop to the phase drop $\varphi$ as  
\be
\Phi = \frac{\Phi_0}{2\pi} \varphi.
\ee
The two-dimensional potential energy of the SQUID has the form \cite{Harris2010}
\ba \label{Ua}
 U(\varphi,\varphi_{\rm CJJ}) =   \frac{(\Phi_0/2\pi)^2}{2 L} (\varphi {-} \varphi^x)^2 \\ 
 + \frac{(\Phi_0/2\pi)^2}{2 L_{\rm CJJ}} (\varphi_{\rm CJJ} {-} \varphi_{\rm CJJ}^x)^2
 - E_J \cos \left(\frac{\varphi_{\rm CJJ}}{2}\right)\cos\varphi.\nonumber
\ea
Phase shifts $\varphi^x$ and  $\varphi_{\rm CJJ}^x$ are proportional to external fluxes $\Phi^x$ and $ \Phi_{\rm CJJ}^{x}$  applied to the main loop and to the CJJ loop, respectively: 
\ba \label{phiX}
 \varphi^x = \frac{2\pi}{\Phi_0} \Big(\Phi^x - \frac{\Phi_0}{2}\Big), \;\;    \varphi_{\rm CJJ}^x = \frac{2\pi}{\Phi_0} \Phi_{\rm CJJ}^{x}.
 \ea
 Here $\Phi_0 {=} h/2e {=} \pi \hbar/e$ is the flux quantum. The Josephson energy in Eq.~\eqref{Ua} is defined as $E_J = (\Phi_0/2\pi) I_C.$

Usually inductance $L_{\rm CJJ}$  is quite small: $L_{\rm CJJ}~\ll~L.$ Therefore, we ignore dynamics of the CJJ loop and assume that $\varphi_{\rm CJJ} \simeq \varphi_{\rm CJJ}^x.$ After these transformations the Hamiltonian \eqref{Ha} of the system can be written as
\ba \label{Hb}
H_{\rm S} = \frac{q^2 }{ 2C} + \frac{(\Phi {-} \Phi^x{+} \Phi_0/2)^2}{ 2L}
\nn 
- E_J\cos \left(\frac{\pi \Phi_{\rm cjj}^x}{\Phi_0} \right) \cos \Big(\frac{2\pi \Phi}{\Phi_0} \Big).
\ea
Flux and charge operators in the Hamiltonian \eqref{Hb} are conjugated, with the commutator $[\Phi, q] = i \hbar.$ 

Let us define current, charge, and voltage operators as
\ba \label{ivA}
I = \frac{\Phi-\Phi^x +\Phi_0/2}{L}, \quad q = - i \hbar\, \frac{\partial }{\partial \Phi}, \quad V = \frac{q}{C},
\ea
respectively.
We also introduce  dimensionless phase and charge operators, $\phi$ and $\hat N$,
\ba \label{phiN}
\phi = \varphi {-} \varphi^x = 2\pi\, \frac{\Phi {-} \Phi^x }{\Phi_0}+\pi, \quad \hat N = - i \frac{\partial}{\partial \phi} = \frac{ q}{2 e}.
\ea
For current and voltage operators we obtain
\ba \label{CurVolt}
I = \frac{\Phi_0}{2 \pi L}\, \phi, \quad V = \frac{2 e}{C}\, \hat N.
\ea
Phase and charge operators $\phi$ and $\hat N$ obey the commutation rule: $ [\phi, \hat N] = i. $
In terms of operators $\phi $ and $N$ the system Hamiltonian \eqref{Hb} has the form
\ba \label{Hc}
H_S = 4 E_C \hat N^2 {+} E_L \frac{\phi^2}{2} {+} E_J \cos \left(\frac{\varphi_{\rm CJJ}^x}{2}\right) \cos \Big(\phi {+} \phi^x \Big), 
\ea
where 
\be
\phi^x = \frac{2 \pi}{\Phi_0} \Phi^x = \varphi^x + \pi
\ee
and $E_C$ and $E_L$ are charging and magnetic energy of the qubit,
\ba
E_C = \frac{e^2}{2 C}, \qquad E_L = \frac{(\Phi_0/2\pi)^2}{L}.
\ea
At $\Phi^x{=}0$ the SQUID is described by a symmetric potential energy.
In the following we choose  $\hbar{=}1$ and $k_B{=}1.$ 

The potential energy of the rf-SQUID has two wells as it is shown in Fig.~\ref{fig1}(b).
 
We partially diagonalize the Hamiltonian $H_S$ in each well and introduce metastable states $\ket{n}$ related to energy levels $E_n$ \cite{QuditPaper}.
Tunneling between state $\ket{m}$ in the left well and state $\ket{n} $ in the right well is provided by a matrix element $\Delta_{mn}$. In the left-right basis the Hamiltonian \eqref{Hb} of the rf-SQUID can be written as
\ba \label{Hd}
H_S = \sum_n E_n \ket{n}\bra{n} - \frac{1}{2} \sum_{m\neq n} \Delta_{mn} \ket{m}\bra{n}.
\ea
This means that
\ba \label{enA}
E_n = \langle n | H_S | n \rangle, \quad \Delta_{m n} = - 2 \langle m | H_S | n \rangle. 
\ea
We note that $\Delta_{mn} = 0$ between two even or two odd states and also that $\Delta_{nm} = \Delta_{mn}^*$.

\section{System-bath interaction}

Here we describe coupling of the dynamical system (rf-SQUID) to a heat bath. The bath has a flux-noise component characterized by  Hamiltonian $H_B^\Phi$ and a charge-noise part, with Hamiltonian $H_B^q.$
The total Hamiltonian $H_B$ of the bath is a sum of these two parts:
\ba \label{bathA}
H_B = H_B^\Phi + H_B^q.
\ea
In addition to the system Hamiltonian \eqref{Hb} and the Hamiltonian \eqref{bathA} describing flux-noise and charge-noise dissipative environment, 
we introduce the Hamiltonian $H_{\rm int}$, which represents interaction of the rf-SQUID with the dissipative environment,
\ba \label{He}
H_{\rm int} = - I \delta\Phi - V \delta q. 
\ea
Here  $I$ is a current in the main loop,  $V = q/C$ is  voltage across CJJ junctions, $q$ is a combined charge of two CJJ junctions, and $C$ is the total capacitance of the CJJ loop. The operator $\delta\Phi$  describes fluctuations of the external flux applied to the main loop of the rf-SQUID, whereas $\delta q$ is the fluctuating charge  on the junction's capacitor.

In terms of dimensionless variables \eqref{phiN}, the  interaction of the SQUID with flux-noise and charge-noise environments is given by 
\ba \label{Hg}
H_{\rm int} = - \phi \, Q_\phi - \hat N\, Q_N
\ea
where $Q_{\phi}$ and $Q_N$ are operators of the flux-noise and charge-noise heat bath,
\ba \label{Qa}
Q_{\phi} = \frac{\Phi_0}{2 \pi L}\, \delta \Phi, \;\; Q_N =\frac{2 e}{C}\, \delta q.
\ea

\subsection{Bath correlators}

We introduce correlation function $K_q(t,t') $ and spectrum $S_q(\omega)$  of charge noise as
\ba \label{korA}
K_q(t,t') = \langle \delta q(t) \delta q(t') \rangle = \int \frac{d\omega}{2\pi} e^{-i\omega (t{-}t')}\, S_q(\omega).
\ea
Flux noise in the main loop of the SQUID is  described by correlator $K_\Phi(t,t')$ and by  spectrum $S_\Phi(\omega)$,
\ba \label{korB}
K_\Phi(t,t') = \langle \delta \Phi(t) \delta\Phi(t') \rangle {=}  \int \frac{d\omega}{2\pi} e^{{-}i \omega (t{-}t')} S_\Phi(\omega).
\ea

\subsection{System-bath Hamiltonian}

In the left-right basis \cite{QuditPaper} introduced in Section \ref{sectionSQUID}, the interaction Hamiltonian has the form
\ba \label{Hf}
H_{\rm int} = -\sum_{m n} (I_{mn} \delta \Phi + V_{mn} \delta q) \ket{m}\bra{n},
\ea
where 
\ba  \label{ivB}
I_{mn} = \bra{m}I\ket{n}, \qquad V_{mn} = \bra{m}V \ket{n}. 
\ea
are matrix elements of current and and voltage of the rf-SQUID. 
Since states $\ket{n}$ are localized in each well, $I_{mn} = 0$ for every pair of states in opposite wells. Also, since $\ket{n}$ is delocalized in charge, we expect $V_{nn} = 0$ for all $n$. Diagonal elements of current are denoted as 
\ba \label{ivC}
I_n \equiv I_{nn} = \bra{n}I\ket{n}.
\ea
States in the different wells have average currents flowing in the opposite directions.

Matrix elements of dimensionless charge, $\hat N,$ and flux, $\phi$, operators in the left-right basis are defined as
\ba \label{yzA}
N_{mn} = \bra{m}\hat N\ket{n}, \qquad \phi_{mn} = \bra{m}\phi\ket{n}, \nn
\phi_n \equiv \phi_{nn} = \bra{n}\phi\ket{n}.\qquad
\ea
The charge operator $N$ has no diagonal matrix elements, $N_{nn}{=}0.$ 
The interaction Hamiltonian \eqref{Hg} has the form
\ba \label{Hh}
H_{\rm int} = - Q_\phi \sum_{mn} \phi_{mn} \, \ket{m}\bra{n} - Q_N \sum_{mn} N_{mn}\, \ket{m}\bra{n},
\ea
with flux and charge-noise operators $Q_\phi$ and $Q_N$ shown in Eqs.~\eqref{Qa}.

In the left-right basis  the Hamiltonian of the rf-SQUID coupled to a dissipative environment  can be written as
\ba \label{Hii}
H = H_0 - \frac{1}{2} \sum_{m\neq n} \Delta_{mn} \ket{m}\bra{n} - Q_{\phi} \sum_{m} \phi_m \ket{m}\bra{m} \nn - Q_{\phi} \sum_{m\neq n} \phi_{mn} \ket{m}\bra{n} - Q_N \sum_{m\neq n} N_{mn} \ket{m}\bra{n} +H_B,
\ea
where $H_B$ is the bath Hamiltonian \eqref{bathA}.
The Hamiltonian
\ba \label{Hk}
H_0 = \sum_n E_n \ket{n}\bra{n}
\ea
presents a contribution of states in left and right wells of the SQUID, and a term with $\Delta_{mn}$  describes tunneling between the wells. We note that $\Delta_{mn}{=}0$ if the states $\ket{m}$ and $\ket{n}$ belong to the same well. 
We also notice that
matrix elements $N_{mn} $ and $\phi_{mn} $ are directly related to charge, $q$, and current, $I$, matrix elements,
\ba \label{yzB}
N_{mn} = \frac{q_{mn}}{2e}, \quad \phi_{mn} = \frac{2 \pi L}{\Phi_0}\, I_{mn}, \; \phi_m = \frac{2 \pi L}{\Phi_0}\, I_{m}.
\ea
In the Hamiltonian \eqref{Hii} we set aside diagonal elements ($\sim \phi_m$) of the system's interaction with the flux-noise bath. A contribution of this part of the system-bath interaction is treated precisely within the hybrid approach \cite{ASMA2018}. In the left-right basis, charge noise is not coupled to any diagonal elements of the system since $N_{mm} {=} 0.$ A contribution of charge noise is considered perturbatively. 

\subsection{Flux noise} 

Flux noise can be described by a dissipative function $f_\Phi(t)$,
\ba \label{fA}
f_\Phi(t) = \int \frac{d \omega}{2\pi} \frac{S_\Phi(\omega)}{\omega^2} \big( 1 - e^{{-}i\omega t} \big).
\ea
Similar functions denoted as $Q_1(t)$ and $Q_2(t)$ have been introduced  \cite{Leggett1987} (see Eqs.~3.36), 
with $Q_1(t) \sim  f_\Phi''(t)$ and with $Q_2(t) \sim f_\Phi'(t)$. 
The flux-noise spectrum 
\ba \label{spectrLH}
S_\Phi(\omega) {=} S_\Phi^L(\omega) {+} S_\Phi^H(\omega)
\ea
has a low-frequency component $S_\Phi^L(\omega)$
and a
high-frequency part $S_\Phi^H(\omega).$ The function $f_\Phi(t)$ also has two components: 
\ba \label{faa}
f_\Phi(t) = f_\Phi^L(t) {+} f^H_\Phi(t),
\ea
each of them is proportional to the corresponding spectral density.
In the Hamiltonian \eqref{Hii} a contribution of diagonal terms, such as 
$Q_{\phi} \sum_{m} \phi_m \ket{m}\bra{m}$, can be treated unperturbatively.
To do that we introduce functions
\ba \label{fB}
f_{mn}(\tau) =  (I_m {-} I_n)^2 f_\Phi(\tau) = f_{mn}^L(\tau) + f_{mn}^H(\tau).
\ea
 Here $I_n {=} \langle n | I | n \rangle {=} \frac{\Phi_0}{ 2\pi L} \phi_n$ is a diagonal matrix element of the current $I$ in the main loop of the SQUID.
 As in Eq.~\eqref{faa}, the dissipative function $f_{mn}(\tau)$ is split into low-frequency and high-frequency parts, with 
\ba \label{fBx}
f_{mn}^{L/H}(\tau){=} (I_m {-} I_n)^2\; f_\Phi^{L/H}(\tau).
\ea
Similar functions can be introduced for charge noise as well. However, there is no need in such functions since the charge operator $q$ has zero diagonal matrix elements in the left-right basis: $q_{nn}{=}\langle n|q |n\rangle {=} 2e N_{nn} {=}0$.

Diagonal elements in the Hamiltonian \eqref{Hii} appear as phase factors, such as $e^{i \tau (I_m-I_n) \delta\Phi }$, in equations for the matrix elements of the system's density matrix. Here $\tau$ is a time step. Averaging these factors over Gaussian fluctuations of the flux $\delta \Phi$ leads to the terms $$\big<e^{i \tau (I_m{-}I_n) \delta\Phi}\big>{=} \exp\Big[ {-}\frac{(I_m{-}I_n)^2 \tau^2}{2} \langle \delta \Phi^2\rangle\Big]{\sim} e^{{-}f_{mn}(\tau)}.$$
For a more precise treatment of these phase factors we introduce the Fourier image \cite{ASMA2018}
\ba \label{gA}
G_{mn}(\omega) = \int d \tau e^{i\omega \tau}\, e^{-f_{mn}(\tau)},
\ea
related to the exponent 
\ba \label{gAx}
e^{-f_{mn}(\tau)} = \int \frac{d\omega}{2\pi}\, e^{- i \omega \tau}\, G_{mn}(\omega).
\ea
The image $G_{mn}(\omega)$  can be represented as a convolution of a low-frequency envelope, $G_{mn}^L(\omega),$ and a high-frequency function
$G_{mn}^H(\omega)$ if we take into account splitting given by Eq.~\eqref{fB}, 
\ba \label{gaa}
G_{mn}(\omega)  = \int \frac{d\Omega}{2\pi} G_{mn}^L(\omega - \Omega) G_{mn}^H(\Omega).
\ea
The low-frequency and high-frequency envelopes in Eq.~\eqref{gaa}  are defined as
\ba \label{gB}
G_{mn}^{L/H}(\omega) = \int_{-\infty}^{+\infty}  d\tau \;e^{i \omega \tau} \, e^{- f_{mn}^{L/H}(\tau)}.
\ea
Functions $G^L_{mn}(\omega), G^H_{mn}(\omega)$  satisfy the normalization condition:
\ba \label{norma}
\int \frac{d\omega}{2\pi} G^{\mu}_{mn}(\omega) = 1,
\ea
where $\mu = L, H.$ The same is true for the function $G_{mn}(\omega).$

\subsubsection{Low-frequency flux noise} 

The function $f_\Phi^L(t)$ is related to the spectrum $S_\Phi^L(\omega).$ This spectrum is peaked at frequencies $\omega$, which are low enough that $|\omega t|\ll 1.$ Therefore, we can use the expansion $$1 - e^{-i\omega t}\simeq i \omega t + \frac{\omega^2 t^2}{2}$$
in the definition \eqref{fA} where $S_\Phi(\omega)$ is replaced by the low-frequency spectrum $S_\Phi^L(\omega).$ 
As a result one obtains a simple formula for $f_\Phi^L(t)$,
\ba \label{fLow}
f_\Phi^L(t) = i \varepsilon_\Phi t + \frac{1}{2} W_\Phi^2 t^2.
\ea
Dissipative parameters $\varepsilon_\Phi$ and $W_\Phi^2$ are determined by the spectrum $S_\Phi^L(\omega)$ of the low-frequency flux noise,
\ba \label{eW}
\varepsilon_\Phi^L = {\cal P} \int \frac{d\omega}{2 \pi} \frac{S_\Phi^L(\omega)}{\omega}, \quad W^2_\Phi = \int \frac{d\omega}{2\pi} S_\Phi^L(\omega).
\ea
Instead of parameters $\varepsilon_\Phi^L$ and $W_\Phi$, we also will use the reorganization energy   $\varepsilon_L$  and the MRT linewidth $W$,  which have a dimension of energy,
 \ba \label{eWL}
\varepsilon_L = \frac{\Phi_0^2}{L^2}\, \varepsilon_\Phi^L, \quad
W = \frac{\Phi_0}{L}\, W_\Phi.
\ea
It follows from the fluctuation-dissipation theorem that $W_\Phi^2 = 2 \varepsilon_\Phi^L T$ and 
 $W^2 = 2 \varepsilon_L T. $ We recall that $\hbar=1$ and $k_B=1$.

The low-frequency component $f_{mn}^L(\tau)$ of the function \eqref{fB}  is determined by the formula
\ba \label{fBy}
f_{mn}^L(\tau) = i \varepsilon_{mn} \tau + \frac{1}{2} W^2_{mn} \tau^2.
\ea
Here,  reorganization energy $\varepsilon_{mn}$ and  MRT linewidth $W_{mn}$ 
 are related to  parameters $\varepsilon_\Phi^L$ and  $W_\Phi$ of the low-frequency flux noise,
\ba \label{eWmn}
\varepsilon_{mn} = (I_m - I_n)^2 \,\varepsilon_\Phi^L, \quad W_{mn} = |I_m - I_n| \,W_\Phi.
\ea
 We notice that $W_{mn}^2 = 2 \varepsilon_{mn} T.$ 
 
A low-frequency envelope defined by Eq.~\eqref{gB} is given by a Gaussian \cite{AminAverin2008},
\ba \label{gC}
G^L_{mn}(\omega) = \sqrt{\frac{2 \pi}{W^2_{mn}}} \exp\Bigg[ - \frac{ ( \omega - \varepsilon_{mn} )^2}{ 2 W^2_{mn}} \Bigg].
\ea

\subsubsection{High-frequency flux noise}

We assume that coupling of the rf-SQUID to high-frequency flux noise is sufficiently weak. 
Therefore, the high-frequency component of the dissipative function \eqref{fB} is small:
$|f_{mn}^H(\tau)|{\ll} 1.$ Expanding the exponential in Eq.~\eqref{gB},  one derives the high-frequency 
function $G_{mn}^H(\omega)$,
\ba \label{gD}
G_{mn}^H(\omega) \simeq 2 \pi \delta(\omega) \Big[ 1 {-} (I_m{-}I_n)^2 \int \frac{d\Omega}{2 \pi} \frac{S_\Phi^H(\Omega)}{\Omega^2} \Big] \nn + 
(I_m{-}I_n)^2 \,\frac{S_\Phi^H(\omega)}{\omega^2}. \;\;
\ea
Approximately, the function \eqref{gD} can be represented as a Lorentzian \cite{ASMA2018}
\ba \label{gE}
G_{mn}^H(\omega) =  \frac{(I_m{-}I_n)^2 S_\Phi^H(\omega)}{ \omega^2 + \Big[ \frac{1}{2} (I_m{-}I_n)^2 S_\Phi(0)\Big]^2 }.
\ea
The first reason for this approximation is that the Lorentzian \eqref{gE}  has the same limit at large frequencies as the expansion \eqref{gD},
\ba \label{gF}
G_{mn}^H(\omega) = (I_m - I_n)^2 \frac{S_\Phi^H(\omega)}{\omega^2}.
\ea
 Secondly, the function \eqref{gE}  satisfies the normalization condition \eqref{norma}. 
 And finally, in the Markovian case, where the spectrum is flat, $S_\Phi^H(\omega){=}S_\Phi^H(0),$   the Lorentzian \eqref{gE} gives the exact expression for the function $ G_{mn}^H(\omega) $. For the Markovian flux-noise bath we have
 \ba \label{fC}
f_{mn}^H(t) = \frac{(I_m-I_n)^2}{2}\, S_\Phi(0)\, |t|.
\ea
With this function the integral in the high-frequency version of Eq.~\eqref{gB} can be calculated precisely
producing the Lorentzian \eqref{gE}.

 In the simplest case flux noise is described by Ohmic spectrum \cite{Leggett1987},
\ba \label{sOhm}
S_\Phi^H(\omega) = \frac{\eta_\Phi\, \omega}{ 1 - e^{-\omega/T}},
\ea
with a coupling constant $\eta_\Phi.$ 
The spectrum \eqref{sOhm} is flat at small $\omega$: $S_\Phi^H(0){=}\eta_\Phi T, $
and linear in $\omega$ at positive  frequencies $\omega \gg T.$ As this takes place, 
the function \eqref{gE} has the form
\ba \label{gOhm}
G_{mn}^H(\omega) = \frac{\eta_{mn}\, \omega}{ 1 - e^{-\omega/T} }\, \frac{1}{ \omega^2 + \big(\frac{\eta_{mn} T}{2} \big) ^2},
\ea
where $\eta_{mn} = (I_m {-}I_n)^2 \eta_\Phi. $

Sub-Ohmic noise \cite{Leggett1987} is characterized by the spectrum
$S_\Phi(\omega) \sim \omega/|\omega|^\alpha$, with a parameter $\alpha >0$ .
This means that for sub-Ohmic noise the ratio $S_\Phi(\omega)/\omega \sim 1/|\omega|^\alpha$ goes down at large positive frequencies, whereas
in the Ohmic case this ratio remains constant: $S_\Phi(\omega)/\omega \sim \eta_\Phi.$

We see from Eq.~\eqref{fA} that the spectrum $S_\Phi(\omega)$ is related to the dissipative function $f_\Phi(t)$.
In its turn, the function $f_\Phi$ is determined by the envelope $G_{mn}(\omega)$ as it follows from Eq.~\eqref{gA}. 
Therefore, in an equivalent manner flux noise can be described by the spectrum $S_\Phi(\omega)$ or by the function $G_{mn}(\omega).$
At large frequencies the correspondence between $S_\Phi^H(\omega)$ and $G_{mn}^H(\omega)$ is illustrated by 
Eq.~\eqref{gF}.

To characterize sub-Ohmic flux noise we introduce the high-frequency function:
\ba \label{gG}
G_{mn}^H(\omega)  = \frac{\kappa}{\tilde \gamma_{mn}} \frac{\beta \omega}{1 - e^{-\beta \omega} }\, \frac{1}{ 1 + (|\omega|/\tilde \gamma_{mn} )^{2 {+} \alpha} },
\ea
where $\beta = 1/T,$ and 
\ba \label{gH}
\tilde \gamma_{mn} = \gamma_\Phi (I_m {-} I_n)^{\frac{2}{1{+}\alpha}}.
\ea
A frequency-independent coefficient $\gamma_\Phi$ defines strength of coupling between the rf-SQUID and high-frequency flux noise,
whereas a parameter $\alpha{\geq} 0$ characterizes a deviation of the flux-noise spectrum from the Ohmic case. 
A factor $\kappa$ in Eq.~\eqref{gG} can be found from the normalization condition \eqref{norma}. 
In the case of weak coupling to high-frequency flux noise, when $\gamma_{mn}\ll T$, the factor $\kappa$ is a function of the sub-Ohmic coefficient $\alpha$ only,
\ba \label{kapA}
\kappa = \pi \frac{2+\alpha}{ \Gamma\Big(\frac{1}{2+\alpha} \Big) \,\Gamma\Big(\frac{1+\alpha}{2+\alpha} \Big)} .
\ea
Ohmic high-frequency flux noise is characterized by zero coefficient $\alpha=0$.

 At large frequencies, $|\omega| \gg \tilde \gamma_{mn},$ we obtain the high-frequency limit of the function \eqref{gF},
\ba \label{gI}
G_{mn}^H(\omega) \simeq \kappa  \frac{\beta \omega}{1 {-} e^{-\beta \omega} }\,\frac{\tilde \gamma_{mn}^{1{+}\alpha}}{|\omega|^{2{+}\alpha}} 
  \simeq  (I_m {-} I_n)^2 \frac{S_\Phi^H(\omega)}{\omega^2}.
\ea
 This means that the envelope \eqref{gG} corresponds to the sub-Ohmic spectrum
\ba \label{gK}
S_\Phi^H(\omega) = \kappa\, \frac{\gamma_\Phi^{1{+}\alpha}}{|\omega|^\alpha} \;\frac{\beta \omega}{1 {-} e^{-\beta \omega} }
\ea
of flux noise.  At $\alpha=0$ the spectrum \eqref{gK} coincides with the Ohmic spectrum \eqref{sOhm} provided that $\gamma_\Phi = \eta_\Phi T/\kappa.$ 
Besides that, the function \eqref{gG} has the same form as Eq.~\eqref{gOhm},
\ba \label{gL}
G_{mn}^H(\omega)  =  \frac{2 \beta \omega}{1 - e^{-\beta \omega} }\, \frac{\tilde \gamma_{mn}}{ \omega^2 + \tilde \gamma_{mn}^2 }.
\ea
Here we take into account that $\kappa{=}2$ at $\alpha{=}0,$ and that $\tilde \gamma_{mn} {=} \eta_{mn} T/2.$

In addition to the coupling strength $\gamma_\Phi$, we can describe high-frequency flux noise in the main loop of the rf-SQUID  by inductive loss tangent $\tan \delta_L(\omega)$ \cite{Nguen2019}, which appears in the expression for the spectrum \eqref{gK},
\ba \label{spBx}
S_\Phi^H(\omega) = 2 L \frac{\tan \delta_L(\omega)}{ 1 - e^{-\omega/T} }.
\ea
Here $L$ is the inductance of the SQUID's main loop. A comparison of Eqs.~\eqref{gK} and \eqref{spBx} leads to the formula for the inductive loss tangent:
\ba \label{tanA}
\tan \delta_L(\omega) = \kappa \frac{\omega}{ 2 L T}\, \frac{ \gamma_\Phi^{1{+}\alpha} }{ |\omega|^\alpha}.
\ea
It is of interest that the coupling strength $\gamma_\Phi$ between the rf-SQUID and high-frequency noise has the following dimension: 
\ba
\gamma_\Phi \sim L^{\frac{1}{1{+}\alpha}} \, |\omega|^{\frac{\alpha}{1{+}\alpha}}.
\ea
For Ohmic flux noise $(\alpha{=}0)$ a parameter $\gamma_\Phi$ has a dimension of inductance, $\gamma_\Phi \sim L.$ 
The linewidth $\tilde \gamma_{mn}$, which appears in Eqs.~\eqref{gG} and \eqref{gH}, has a dimension of energy at any sub-Ohmic coefficient $\alpha$,
\ba
\tilde \gamma_{mn} \sim \Big[ \frac{ L (I_m{-}I_n)^2}{\omega } \Big]^{\frac{1}{1{+}\alpha}}\, \omega.
\ea

Flux noise also can be characterized by shunting resistance $R_s$, which is inversely proportional to the loss tangent $\tan \delta_L(\omega)$,
\ba \label{resA}
\frac{1}{R_s(\omega)} = \frac{\tan \delta_L(\omega)}{\omega L}.
\ea
It follows from Eq.~\eqref{tanA} that
\ba \label{resB}
\frac{1}{R_s(\omega)} = \frac{\kappa}{2 L^2 T} \, \frac{\gamma_\Phi^{1{+}\alpha}}{|\omega|^{\alpha} }.
\ea
Instead of the coupling strength $\gamma_\Phi$ we can introduce inductance $L_\Phi(\omega)$,
\ba \label{La}
L_\Phi(\omega) = \gamma_\Phi \Big( \frac{\gamma_\Phi}{ |\omega|} \Big)^\alpha.
\ea
The spectrum of high-frequency flux noise \eqref{gK} is proportional to inductance $L_\Phi(\omega)$,
\ba \label{spC}
S_\Phi^H(\omega) = \kappa\, L_\Phi(\omega) \;\frac{\beta \omega}{1 {-} e^{-\beta \omega} }.
\ea
Inductance $L_\Phi(\omega)$ is inversely proportional to shunting resistance, 
\ba \label{resC}
\frac{1}{R_s(\omega)} = \frac{\kappa}{2 L T} \, \frac{L_\Phi(\omega)}{L}.
\ea
The spectrum $S_\Phi^H(\omega)$ can be represented as
\ba \label{spD}
S_\Phi^H(\omega) = 2 \frac{L^2}{R_s(\omega)} \, \frac{\omega}{ 1 - e^{-\omega/T} }.
\ea
For Ohmic flux noise, where $\alpha=0,$ dissipative parameters $R_s$ and $ L_\Phi$ do not depend on frequency $\omega.$

Coupling to high-frequency flux noise also can be characterized by a parameter 
\ba \label{gam0}
\lambda_\Phi = \gamma_\Phi \Bigg( \frac{\Phi_0}{L} \Bigg)^\frac{2}{1{+}\alpha},
\ea
which has a dimension of energy. The high-frequency envelope $G^H_{mn}(\omega)$ given by Eq.~\eqref{gG} is determined by the parameter $\tilde \gamma_{mn}$, 
\ba \label{gam00}
\tilde \gamma_{mn} = \lambda_\Phi\, \Bigg[ \frac{L (I_m - I_n) }{\Phi_0} \Bigg]^\frac{2}{1+\alpha}.
\ea
Both parameters, $\tilde \gamma_{mn}$ and $\lambda_\Phi$, have a dimension of energy. However, $\lambda_\Phi$ does not depend on the states $\ket{n}$ and $\ket{m}$ of the rf-SQUID nor on the non-Ohmic parameter $\alpha.$

\subsection{Charge noise}

The spectrum of charge noise, $S_q(\omega),$ is defined by Eq.~\eqref{korA}.  This spectrum can be described by the formula
\ba \label{korE}
S_q(\omega) = 2 C\, \frac{\tan \delta_C(\omega)}{ 1 - e^{-\omega/T} },
\ea
where $\tan \delta_C(\omega)$ is a frequency-dependent loss tangent of charge noise \cite{Nguen2019}. 
The spectral density $S_q(\omega)$ should be flat at $\omega = 0$. To satisfy this condition we assume that
\ba \label{tanC}
\tan \delta_C(\omega) = \tan \delta_C\, \tanh\Big(\frac{\omega}{T}\Big),
\ea
where $\tan \delta_C$ is a constant characterizing the strength of coupling between the rf-SQUID qubit and charge-noise dissipative environment.

\section{Macroscopic resonant tunneling}
\label{sectionMRT}

The MRT  experiment \cite{HarrisMRT2008,LantingMRT2011,AW2022}  is performed by initializing the rf-SQUID in the ground state $\ket{0}$ of the left well with  probability $P_0 = 1.$ We are interested in the escape rate $\Gamma_0$ from the initial state to the states in the right well. This rate is measured as a function of the external flux bias $\Phi^x$ applied to the main loop of the rf-SQUID. 
The escape rate $\Gamma_0(\Phi^x)$  is given by 
\ba \label{gamA}
\Gamma_0(\Phi^x) = - \Bigg[ \frac{ d P_0}{ d t} \Bigg]_{t{=}0} = \sum_{{\rm odd}\; n} \Gamma_{0 n} (\omega_{0 n}),
\ea
where $P_0(t)$ is the probability to be in the state $\ket{0}$ at the moment of time $t$, and 
$ \Gamma_{0 n} (\omega_{0 n})$ is the escape rate from  $\ket{0}$ to state $\ket{n}$ in the right well. 
This rate depends on energy separation $\omega_{0 n}{=} E_0{-}E_n$ between  the initial and target states. The energy distance $\omega_{0 n}$ can be tuned by changing the external flux $\Phi^x.$ 
The  MRT experiment can be repeated in the opposite direction when the system is initially localized in the ground state $\ket{1}$ of the right well and tunneling happens from the right well to the left well. We assume that the tunneling matrix element $\Delta_{0n}$ is much less than the distance between energy levels in each well.

The rate $\Gamma_{0 n} (\omega_{0 n} ) $ peaks  when $\omega_{0 n} \simeq 0. $ At these resonant conditions the ground-state energy level $\ket{0}$ in the left well is aligned with a level $\ket{n}$ in the right well. The specific shape of the function $\Gamma_{0 n} (\omega) $  is determined by details of flux and charge noise acting on the rf-SQUID.
Whereas flux noise affects the transition peak directly, charge noise broadens the peak indirectly via intrawell relaxation.  When state $\ket{n}$ is an excited state in the target (right) well, the system quickly relaxes down to the lowest energy state, $\ket{1}$, within the well. This intrawell relaxation is typically dominated by charge noise and provides an additional broadening to the transition peak.

\subsection{Master equations}

A time evolution of the probability distribution $P_n(t)$ over the states of rf-SQUID is governed by the  master equation \eqref{Mt2} derived in Appendix~A, 
\ba \label{mqA1}
\dot P_n + \sum_m \Gamma_{mn} P_n   =  
\sum_m \Gamma_{nm} P_m.
\ea
The relaxation matrix $\Gamma_{mn}$ is represented as a convolution of the Lorentzian, which  describes broadening of energy levels participating in the transition, times the hybrid spectrum $S_{nm}(\omega{-}\omega_{mn})$ of the bath,
\begin{widetext}
\ba \label{mqA2}
\Gamma_{mn} = 2 \int \frac{d\omega}{2\pi} \frac{  \gamma_n(\omega) + \gamma_m(-\omega)}{[\omega{-}\delta_n(\omega){+}\delta_m({-}\omega)]^2
+ [\gamma_n(\omega) {+}\gamma_m(-\omega)]^2 }\, S_{nm}(\omega-\omega_{mn}).
\ea
\end{widetext}
The hybrid-noise spectrum $S_{mn}(\omega)$ itself
 is equal to a convolution of the low-frequency Gaussian $G_{mn}^L(\omega)$ given by Eq.~\eqref{gC} multiplied by a combination of the high-frequency Lorentzian $G_{mn}^H(\omega)$ and the charge-noise spectrum $S_q(\omega),$
\ba \label{spectrA}
S_{mn}(\omega) = \int \frac{d\Omega}{2\pi} G_{mn}^L(\omega{-}\Omega) \times \nn \Big[ |\tilde \Delta_{mn}(\omega) |^2  G_{mn}^H(\Omega) + \Big|\frac{q_{mn}}{C}\Big|^2 S_q(\Omega)
 \Big],
\ea
It should be mentioned that $S_{nn}(\omega)=0$ by definition.
In Eq.~\eqref{spectrA} we introduce a frequency-dependent tunneling amplitude between states $\ket{m}$ and $\ket{n}$ ($m\neq n$):
\ba \label{dL}
\tilde \Delta_{mn}(\omega) = \frac{1}{2}\, \Delta_{mn} + \omega \frac{I_{mn}}{I_m - I_n},
\ea
where $\Delta_{mn}$ is 
the tunneling amplitude   given by Eq.~\eqref{enA}, and $I_{mn}$ are matrix elements of the current operator defined by Eqs.~\eqref{ivB}. 

The Lorentzian in Eq.~\eqref{mqA2} is characterized by the combined line width $\gamma_n(\omega){+}\gamma_m(-\omega)$ 
and by the frequency shift $\delta_n(\omega){-}\delta_m(-\omega)$.
The broadening $\gamma_n$  and the line shift $\delta_n$ of the $n$th level,
\ba \label{gammaN}
 \gamma_n(\omega) =  \frac{1}{2} \sum_k S_{nk}(\omega + \omega_{nk}), \\
\label{shiftN}
 \delta_n(\omega) = \nn \sum_k \int_0^\infty \frac{d\Omega}{2\pi} \, \frac{ S_{nk}(\omega{+}\omega_{nk}{-}\Omega) - S_{nk}(\omega{+}\omega_{nk}{+}\Omega)}{\Omega},
 \ea
 are shown in Eqs.~\eqref{gamXa}.

\subsection{Escape rate}

The master equation \eqref{mqA1} governs dissipative dynamics of the rf-SQUID in the regime of macroscopic resonant tunneling.
In order to describe the initial stage of the transition from the left-well ground state $\ket{n}=\ket{0}$ to the states in the right well we have to assume in Eq.~\eqref{mqA1} that $P_n(0)=0$ for all states $\ket{n}$ except the initial state $\ket{0},$ which has the probability $P_0=1.$ This state has zero linewidth: $\gamma_0=0.$ We also omit insignificant frequency shifts $\delta_0$ and $\delta_m$ of energy levels in the left and right wells.The escape rate $\Gamma_0{=}\sum_m \Gamma_{m0}$ from the state $\ket{0}$ to states in the right well is determined by the  convolution of the Lorentzian describing lineshape of the target level $\ket{m}$ and the hybrid-noise spectrum 
$S_{0m}(\omega_{0m}{-}\omega)$,
\ba \label{Esc1}
\Gamma_0 = \sum_{m\neq 0}  \int \frac{d\omega}{2\pi}  \frac{ 2 \gamma_m(\omega) }{ \omega^2
+ \gamma_m^2(\omega)}    S_{0m}(\omega_{0m}{-}\omega).
\ea
Here, the hybrid spectrum $S_{0n}(\omega)$ is also represented by a convolution given by Eq.~\eqref{spectrA}. 

Therefore, the MRT rate $\Gamma_0$ is described by a double convolution 
\ba \label{gammaL}
\Gamma_0 = \sum_{m=1,3,5,..} \int \frac{d \omega_1}{2 \pi} \int \frac{d\omega_2}{2\pi} \times \nn \frac{2 \gamma_m(\omega_1)}{ \omega_1^2 + \gamma_m^2(\omega_1)} \,  G_{0m}^L(\omega_{0n}  {-}\omega_1{-}\omega_2) \times \nn  \Big[ \Big|\frac{\Delta_{0m}}{2}\Big|^2 G_{0 m}^H(\omega_2) + 2 \frac{|q_{0m}|^2}{C} \, \frac{\tan \delta_C(\omega_2)}{ 1 - e^{-\omega_2/T} } \Big].
\ea
Here we take into account that in Eq.~\eqref{gammaL}  the current matrix elements $I_{0m}=0$ since the states $\ket{0}$ and $\ket{m}$  belong to the opposite wells. 

The line broadening $\gamma_m(\omega)$ of the $m$-th level has contributions of high-frequency flux noise, $\gamma_m^F,$ and high-frequency charge noise, $\gamma_m^C$,
\ba \label{gamI}
 \gamma_m(\omega) = \gamma_m^F(\omega) + \gamma_m^C(\omega),
\ea
where
\ba\label{gamK}
\gamma_m^F(\omega) &=& 
\frac{1}{2}\sum_{k=1,3,..} \frac{|I_{mk}|^2}{(I_n{-}I_k)^2} \, \omega^2\, G^H_{mk}(\omega), 
 \nn 
\gamma_m^C(\omega) &=& 
\sum_{k=1,3,..} \frac{|q_{mk}|^2}{C} \frac{\tan \delta_C(\omega)}{ 1 - e^{-\omega/T} }.
\ea
For the state $\ket{m}$ the rates \eqref{gamK} can be reduced to
\ba\label{gamKa}
\gamma_m^F(\omega) &=& 
\frac{1}{2}\sum_{k=1,3,..} |I_{mk}|^2 S_\Phi(\omega), \nn 
\gamma_m^C(\omega) &=& 
\frac{1}{2}\sum_{k=1,3,..}  |V_{mk}|^2  S_q(\omega),
\ea
which represent the standard single photon intrawell relaxation channels.

The low-frequency Gaussian $G_{0m}^L(\omega)$ and the high-frequency Lorentzian $G_{0m}^H(\omega)$ are given by Eqs.~\eqref{gC}
and \eqref{gG} in the general case and by Eq.~\eqref{gL} in the Ohmic flux noise case. 
We note that in the left-right basis \cite{QuditPaper}  
the matrix $\Delta_{mn}$ has zero intrawell matrix elements and non-zero interwell elements. Contrary to this, the current's  matrix $ I_{mn}$ has zero interwell matrix elements and non-zero intrawell elements. The charge matrix $q_{mn}$ has both intrawell and interwell elements although intrawell elements are bigger than interwell elements. 

The last line in the rate \eqref{gammaL} represents interwell relaxation due to flux and charge noise. In flux qubits with a large tunneling barrier, the contribution of charge noise to interwell tunneling is small. One can therefore write
\ba \label{gammaLx}
\Gamma_0 = \sum_{m=1,3,5,..} \Big|\frac{\Delta_{0m}}{2}\Big|^2 \int \frac{d \omega_1}{2 \pi} \int \frac{d\omega_2}{2\pi} \times \nn \frac{2 \gamma_m(\omega_1)}{ \omega_1^2 + \gamma_m^2(\omega_1)} \,  G_{0m}^L(\omega_{0n}  {-}\omega_1{-}\omega_2) G_{0 m}^H(\omega_2).
\ea
This is the equation that is used to fit with experimental data in Ref.~\cite{AW2022}.

\begin{figure}
\includegraphics[width=\columnwidth]{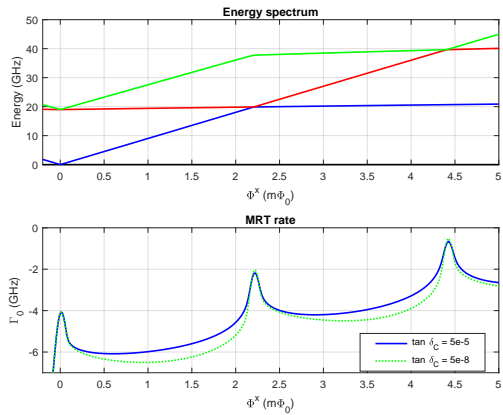}
\caption{\label{ThreePeaks} MRT signal for two values of coupling between the rf-SQUID and charge noise: 1) $\tan \delta_C = 5\times 10^{-3}$ (blue continuous line; 2) $\tan \delta_C = 5\times 10^{-8}$ (green dotted line).}
\end{figure}

\subsection{Example}

As an example we analyze  macroscopic quantum tunneling in the rf-SQUID having  parameters shown in Table~\ref{tab:table1}.
\begin{table}[h!]
\caption{Parameters of the rf-SQUID}
    \label{tab:table1}
\begin{center}
\begin{tabular}{ c|c } 
 \hline
 L & 250~pH  \\ 
 $L_{\rm CJJ}$ & 14~pH  \\ 
 C & 110~pH  \\ 
 $I_C$ & 2.3~$\mu$A\\
 \hline
\end{tabular}
\end{center}
\end{table}
Characteristics of flux and charge noise are shown in Table~\ref{tab:table2}.
\begin{table}[h!]
  \begin{center}
    \caption{Dissipative characteristics}
    \label{tab:table2}
    \begin{tabular}{c|c} 
    \hline
      T & 10~mK\\
      W & 28~mK\\
      $\lambda_\Phi$ & 9.6~mK\\
      $\alpha$ & 0 \\
      \hline
    \end{tabular}
  \end{center}
\end{table}
In Fig.~\ref{ThreePeaks} we see three MRT peaks in the range of external fluxes  $-0.2~m\Phi_0 \leq \Phi^x \leq 5~m\Phi_0.$
Each peak corresponds to the alignment of the ground energy level in the initial (left) well with: 

\noindent
(0) the ground energy level in the target well (zero-order peak at $\Phi^x\simeq 0$); 

\noindent
(1) the first excited state in the target well (first-order peak at $\Phi^x \simeq 2.2~{\rm m}\Phi_0$);

\noindent
(2) the second excited state in the target well (second-order peak at $\Phi^x \simeq 4.4~{\rm m}\Phi_0$).

\noindent
In Fig.~\ref{ThreePeaks}  we show the MRT rate for two values of coupling to charge noise characterized by  the charge-noise loss tangent $\tan \delta_C = 5\times 10^{-3}$ (blue continuous line) and by 
$\tan \delta_C = 5\times 10^{-8}$ (green dotted line). Evidently, the charge noise has no effects on the peaks themselves. However, charge fluctuations have a significant impact on the MRT signal in valleys between peaks.

\section*{Conclusion}

We have analyzed multilevel macroscopic quantum tunneling in the rf-SQUID flux qubit. Within the hybrid-noise approach to the theory of open quantum systems we have obtained the set of master equations \eqref{mqA1} for the probability distribution $P_n(t)$ over states of the rf-SQUID.
The multilevel MRT signal is determined by the rate $\Gamma_0$ of escape from the left-well ground state to states in the right well. We have shown that this rate is represented as a double convolution \eqref{gammaL} of a Lorentzian describing charge-noise broadening of right-well levels multiplied by low-frequency flux noise Gaussian times a sum of high-frequency flux-noise Lorentzian plus charge-noise spectrum. The multilevel MRT rate $\Gamma_0$ measured as a function of the external flux $\Phi^x$ demonstrates many peaks, each related to the resonance between the initial state in the left well and some of the states in the target well. These peaks and the valleys between them
contain information about both flux and charge sources of noise as presented in Ref.~\cite{AW2022} Flux noise mainly defines the shape of the MRT peaks, whereas the effects of charge noise are especially visible in inter-peak valleys.  

\section{Acknowledgements}

We are grateful to Alex Sato and Mark Johnson for careful reading of  the manuscript.

\appendix

\section{Derivation of master equations}

Here we derive a set of master equations describing incoherent tunneling between left and right wells in the presence of flux and charge noise. We follow a hybrid approach to the theory of open quantum systems developed in Ref.~\cite{ASMA2018}. Within this approach an interaction of the system (rf-SQUID) with the bath described by diagonal elements of the total Hamiltonian is treated precisely, whereas off-diagonal elements of the Hamiltonian are considered as a perturbation. Other ways to derive master equations  for open quantum systems can be found in Refs.~\cite{Lidar2013,Lidar2020}.
It should be emphasized that a perturbative
treatment of a system-bath interaction does not work for modern quantum annealers as it is shown in Ref.~\cite{Lidar2022}. The same is true for many problems of quantum biology \cite{Lambert2013}. Here we could mention Refs.~\cite{Kubo1989,Aki2005,Aki2009,Ghosh2011} where authors have tried to go beyond the perturbation theory.

\subsection{ Schr\"odinger equation}

In the left-right basis and in the Schr\"odinger representation  the system-bath Hamiltonian has the form
\ba \label{Hi}
H_{SB} = \sum_n (E_n - I_n \delta \Phi )\ket{n}\bra{n} \nn - \sum_{m\neq n} \Big( \frac{\Delta_{mn}}{2} + I_{mn} \delta \Phi + V_{mn} \delta q \Big)\ket{m}\bra{n}  + H_B.
\ea
Here $I_{mn}$ and $V_{mn}$ are matrix elements \eqref{ivB} of current and voltage, with $I_n \equiv I_{nn},$ and  
\ba \label{HbA}
H_B = H_B^{\Phi} + H_B^q
\ea
 is the Hamiltonian of the flux-noise and charge noise heat bath.
A time evolution of the system-bath wave function is governed by the Schr\"odinger equation
\ba \label{Sch1}
i \frac{\partial}{\partial t } \ket{\Psi_{SB}(t)} = H_{SB} \ket{\Psi_{SB}(t)}.
\ea
A formal solution of this equation can be written as
\ba \label{Sch1a}
 \ket{\Psi_{SB}(t) } = U(t) \ket{\Psi_{SB}(0)},
 \ea
with the unitary matrix
\ba \label{Sch1b}
U(t) = e^{-i H_{SB} t}.
\ea
In the interaction picture of quantum mechanics the solution of the 
 Schr\"odinger equation \eqref{Sch1} can be found in a few steps. 
We begin with removing the bath Hamiltonian from the Hamiltonian \eqref{Hi}. To do this we assume that
\ba \label{Sch2}
\ket{\Psi_{SB}(t)} = U_B(t) \ket{\Psi(t)}.
\ea
Here the unitary matrix 
\ba \label{SchUB}
U_B(t) = e^{- i H_B t}
\ea
determines the free evolution of bath variables. The function $\ket{\Psi(t)}$ obeys the equation
\ba \label{Sch3}
i \frac{\partial}{\partial t } \ket{\Psi(t)} = H(t) \ket{\Psi(t)},
\ea
with the Hamiltonian
\ba \label{Sch4}
H(t) = U_B^\dag H_{SB} U_B - i U_B^\dag \dot U_B.
\ea
Taking into account Eq.~\eqref{SchUB} we obtain
\ba \label{Sch5}
H(t) = \sum_n [ E_n - I_n \delta \Phi(t) ] \ket{n}\bra{n}  \nn - \sum_{m\neq n} \Big[ \frac{\Delta_{mn}}{2} + I_{mn} \delta \Phi(t) + V_{mn} \delta q(t) \Big]\ket{m}\bra{n}. 
\ea
Here
\ba \label{SchBath}
 \delta \Phi(t) = U_B^\dag(t)\, \delta \Phi\, U_B(t), \;
 \delta q(t) = U_B^\dag(t)\, \delta q\, U_B(t) \;
 \ea
 are free-evolving flux-noise and charge-noise bath operators.
 
 Our next goal is to remove diagonal terms from the Hamiltonian \eqref{Sch5}. It can be done with the unitary matrix
 \ba \label{Udiag}
 U_{\rm diag}(t) = \sum_n  e^{-i E_n t}\, U_n(t) \ket{n}\bra{n},
 \ea
 where
 \ba \label{UnMat}
  U_n(t) = \mathcal T \exp \Big\{ i \int_0^t d\tau I_n \delta \Phi(\tau)  \Big\},
 \ea
 and $\mathcal T$ denotes time ordering of subsequent operators.  
 The solution of Eq.~\eqref{Sch3} can be written as
 \ba \label{Sch6}
 \ket{\Psi(t)} = U_{\rm diag}(t) \ket{\tilde \Psi(t)}. 
 \ea
 The state $\ket{\tilde \Psi(t)}$ obeys the equation
 \ba \label{Sch7}
 i \frac{\partial}{\partial t } \ket{\tilde \Psi(t)} = \tilde H(t) \ket{\tilde \Psi(t)}
 \ea
 with the Hamiltonian
 \ba \label{Sch8}
 \tilde H(t) = U_{\rm diag}^\dag\, H(t) \,U_{\rm diag}  - i U_{\rm diag}^\dag\, \dot U_{\rm diag}.
 \ea
 Considering Eqs.~\eqref{Udiag} and \eqref{UnMat} we find that the Hamiltonian $\tilde H(t)$ can be written as
 \ba \label{Sch9}
 \tilde H (t) = - \sum_{k\neq l} \tilde Q_{k l}(t)\, \ket{k}\bra{l},
 \ea
 with
 \ba \label{Bath1}
 \tilde Q_{k l}(t) =   e^{i \omega_{kl} t}\; Q_{kl}(t).
 \ea
 Here 
 \ba \label{Bath2}
 Q_{kl}(t) =\nn  
  U_k^\dag(t) \Big[ \frac{\Delta_{mn}}{2} {+} I_{mn} \delta \Phi(t) {+} V_{mn} \delta q(t) \Big] U_l(t).
 \ea
 is the bath operator, and
 $\omega_{k l} = E_k - E_l$ is the energy spectrum.
 
 The solution of Eq.~\eqref{Sch7},
 \ba \label{Sch10}
 \ket{\tilde \Psi(t)} = \tilde U(t) \ket{\tilde \Psi(0)},
 \ea
 is determined by the unitary matrix
 \ba \label{Sch11}
 \tilde U(t) = \mathcal T \exp\Big[ - i \int_0^t d\tau \,\tilde H(\tau) \Big].
 \ea
 This matrix is a functional of free bath operators $\tilde Q_{kl}(t),$
 \ba \label{Sch12}
 \tilde U(t) = \mathcal T \exp\Big[  i \sum_{kl} \int_0^t d\tau \,\tilde Q_{kl}(\tau) \, \ket{k(\tau)}\bra{l(\tau)} \Big].
 \ea
 Here we use Eq.~\eqref{Sch9} and assume for a moment that the system's basis states $\ket{k(\tau)}, \ket{l(\tau)}$
 depend on time as it happens, for example, for the instantaneous basis in quantum annealing. \cite{ASMA2018}
 
Combining Eqs.~\eqref{Sch2}, \eqref{Sch6}, and \eqref{Sch10} we find another representation of the system-bath evolution matrix \eqref{Sch1b}:
\ba \label{Sch14}
U(t) = U_B(t)\, U_{\rm diag}(t)\, \tilde U(t).
\ea

\subsection{ Density matrix of the system}

The system-bath density matrix $\rho_{SB}(t)$ is defined in terms of the state \eqref{Sch1a},
\ba \label{Sch15}
\rho_{SB}(t) = \nn  \ket{\Psi_{SB}(t)}\bra{\Psi_{SB}(t)} = U(t) \rho_{SB}(0) U^\dag(t),
\ea
with the initial condition $\rho_{SB}(0) = \ket{\Psi_{SB}(0)}\bra{\Psi_{SB}(0)}.$
The system's density matrix $\rho_S$ is defined as a trace of the total matrix over bath degrees of freedom,
\ba \label{Sch16}
\rho_S(t) = \textrm{Tr}_B \rho_{SB}(t) = \sum_{m n} \rho_{nm} \, \ket{n}\bra{m}.
\ea
Matrix elements of the system density operator in the left-right basis are defined as
\ba \label{Sch17}
\rho_{nm}(t) = \langle n |\rho_S(t) |m \rangle = \textrm{Tr}_B \langle n |\rho_{SB}(t)|m\rangle.
\ea
Taking into account Eq.~\eqref{Sch15} we obtain
\ba \label{Sch18}
\rho_{nm}(t) = {\rm Tr}_B \langle n | U(t) \rho_{SB}(0) U^\dag(t) |m\rangle =\nn  {\rm Tr}_B \sum_k \langle n| U(t) |k\rangle \langle k|\rho_{SB}(0) U^\dag(t) |m\rangle  = \nn \sum_k {\rm Tr}_B \langle k| \rho_{SB}(0) U^\dag(t) \ket{m}\bra{n} U(t) \ket{k}.
\ea
Here we resort to the completeness condition: $\sum_k \ket{k}\bra{k} = 1$. A trace over system states is defined as $${\rm Tr}_S (\ldots) = \sum_k \langle k|(\ldots )|k\rangle.$$ Using this definition we find that the matrix elements of the system's density operator are represented by a trace
${\rm Tr} = {\rm Tr}_B {\rm Tr}_S$ over both system and bath degrees of freedom:
\ba \label{Sch19}
\rho_{nm}(t) = {\rm Tr} [ \rho_{SB}(0) U^\dag(t) \ket{m}\bra{n} U(t)].
\ea
The unitary matrix $U(t)$ is given by Eq.~\eqref{Sch14}, therefore,
\ba \label{Sch20}
\rho_{nm}(t) = \nn e^{i\omega_{mn} t} \, {\rm Tr} [ \rho_{SB}(0) \tilde U^\dag(t) U_m^\dag(t) \ket{m} \bra{n} U_n(t) \tilde U(t)].\;
\ea
We are interested in a time evolution of a probability $P_n(t) $ to find the system (rf-SQUID) in the state $\ket{n}.$ This probability is determined by the diagonal elements of the system density matrix: 
\ba \label{Sch21}
P_n(t) = \rho_{nn}(t) = {\rm Tr} [\rho_{SB}(0) \tilde U^\dag(t)  \ket{n} \bra{n}\tilde U(t)].
\ea
Assuming that at the beginning of evolution the system-bath density matrix is factorized via $\rho_{SB}(0) = \rho_S(0)\otimes \rho_B(0)$, we obtain
\ba \label{Sch22}
P_n(t) = {\rm Tr}_S\Big[ \rho_S(0) \Big< \tilde U^\dag(t) \ket{n}\bra{n} \tilde U(t) \Big>\Big].
\ea
Here a notation $\langle \ldots \rangle$ means averaging over bath degrees of freedom,
\ba \label{BathAv}
\langle  \mathcal A \rangle = {\rm Tr}_B [ \rho_B(0) \mathcal A],
\ea
with $\mathcal A $ being an arbitrary system-bath operator.

It is convenient to introduce a Heisenberg operator 
\ba \label{sigMN}
\sigma_{mn}(t) = \tilde U^\dag(t) \ket{m}\bra{n} \tilde U(t),
\ea
which depends on the unitary matrix $\tilde U(t)$ only. We see from Eq.~\eqref{Sch22} that the probability $P_n$ is determined by the operator
\ba \label{sigN}
\sigma_n(t) \equiv \sigma_{nn}(t) = \tilde U^\dag(t) \ket{n}\bra{n} \tilde U(t)
\ea
averaged over system and bath initial states:
\ba \label{probSig}
P_n(t) = {\rm Tr}_S [ \rho_S(0) \langle \sigma_n(t) \rangle ].
\ea

\subsection{Heisenberg equation}

A time evolution of the probability $P_n(t)$ to find the system in the state $\ket{n}$ is determined by the operator $\sigma_n(t)$ as it follows from Eq.~\eqref{probSig}. Here we derive  a  Heisenberg equation for a more general operator $\sigma_{mn}(t)$ defined by Eq.~\eqref{sigMN}.
To do this we take a time derivative of the operator \eqref{sigMN}:
\ba \label{LanA}
\dot \sigma_{mn} = i \,\tilde U^\dag\, [\tilde H, \ket{m}\bra{n}]\, \tilde U.
\ea
We consider that $\dot{\tilde U} = - i\, \tilde H\, \tilde U, $ as it is evident from Eq.~\eqref{Sch11}.  Using Eq.~\eqref{Sch9} one calculates a commutator in Eq.~\eqref{LanA} and obtains a Heisenberg equation for the operator \eqref{sigMN}:
\ba \label{LanB}
\dot \sigma_{mn} = i \sum_k \Big[ \tilde U^\dag \tilde Q_{nk} \ket{m}\bra{k} \tilde U -
 \tilde U^\dag \tilde Q_{km} \ket{k}\bra{n} \tilde U \Big].\;
 \ea
 All operators in Eq.~\eqref{LanB} are taken at the moment of time $t$. 
 
 \subsection{Bath correlators}
 
 The Heisenberg equation \eqref{LanB} includes free-evolving bath operators $\tilde Q_{nk}(t)$ and $\tilde Q_{km}(t)$ defined in Eq.~\eqref{Bath1}. These operators have a linear dependence on charge-noise variable $\delta q$. At the same time they are non-linear functionals  of flux noise operators $\delta \Phi$ as it follows from Eqs.~\eqref{UnMat}, \eqref{Bath1}, and \eqref{Bath2}. 
 
 We assume that non-diagonal system-bath coupling described by the Hamiltonian \eqref{Sch9} is weak. Therefore, a time evolution of operators $\sigma_{mn}$ is mainly determined by the second-order correlation function $\tilde K_{mn}^{n'm'}(t,t') $ of bath operators $\tilde Q_{mn}(t)$ and $\tilde Q_{n'm'}(t'):$
 \ba \label{LanKa}
 \tilde K_{mn}^{n'm'}(t,t') = \langle \tilde Q_{mn}(t) \tilde Q_{n'm'}(t')\rangle.
 \ea
 We notice that $\langle \tilde Q_{mn}(t)\rangle = 0.$ According to Eq.~\eqref{Bath1} the bath variable $\tilde Q_{mn}(t)$ is related to the variable $Q_{mn}(t)$ defined in Eq.~\eqref{Bath2}. Correlators of these two variables, $\tilde K_{mn}^{n'm'}(t,t')$ and 
 \ba \label{LanK1}
 K_{mn}^{n'm'}(t,t') = \langle Q_{mn}(t)  Q_{n'm'}(t')\rangle
 \ea
 are related to each other,
 \ba \label{LanK2}
 \tilde K_{mn}^{n'm'}(t,t') = e^{i\omega_{mn}t}\, e^{i\omega_{n'm'}t'} K_{mn}^{n'm'}(t,t').
 \ea
 It is shown \cite{ASMA2018} that
 components with $m'=m$ and $n'=n$ dominate in the set of correlators \eqref{LanK1}, therefore,
 \ba \label{LanKb}
 K_{mn}^{n'm'}(t,t')  = \delta_{m m'} \delta_{n n'}\,  K_{mn}(t,t'),
 \ea
 where 
 \ba \label{LanX1}
 K_{mn}(t,t') {=} \langle Q_{mn}(t) Q_{nm}(t')\rangle = \nn \int \frac{d\omega}{2 \pi} \, e^{-i\omega(t-t')}\,  S_{mn}(\omega).
 \ea 
 This correlator is related to the spectrum $S_{mn}(\omega).$
 Similar to Eq.~\eqref{LanKb}, the correlator $\tilde K_{mn}^{n'm'}(t,t')$ is determined by the function 
 \ba \label{LanK3}
 \tilde K_{mn}(t,t') = \langle \tilde Q_{mn}(t) \tilde Q_{nm}(t') \rangle =\nn e^{i\omega_{mn}(t{-}t')} K_{mn}(t,t').
 \ea
 The correlator $\tilde K_{mn}(t,t')$ is defined by the spectral function $\tilde S_{mn}(\omega), $
 \ba \label{LanX2}
 \tilde K_{mn}(t,t') = \int \frac{d\omega}{2 \pi} \, e^{-i\omega(t-t')}\,  \tilde S_{mn}(\omega).
\ea
 It follows from Eq.~\eqref{LanK3} that
 \ba \label{LanK5}
 \tilde S_{mn}(\omega) = S_{mn}(\omega + \omega_{mn}).
 \ea
 The spectrum $S_{mn}(\omega) $ can be calculated using Eqs. \eqref{Bath2} and \eqref{LanX1} within the context of the hybrid-noise approach, \cite{ASMA2018}
\ba \label{spAx}
S_{mn}(\omega) = \int \frac{d\Omega}{2\pi} G_{mn}^L(\omega-\Omega) \times \nn
\Big[ |\tilde \Delta_{mn}(\omega)|^2\, G_{mn}^H(\omega) + \frac{|q_{mn}|^2}{C^2} S_q(\Omega) \Big],
\ea
where
\ba \label{spAy}
\tilde \Delta _{mn}(\omega) = \frac{\Delta_{mn}}{2} + \omega \frac{I_{mn}}{I_m - I_n}
\ea
is a frequency-dependent tunneling rate, 
$ G_{mn}^L(\omega) $ is a low-frequency envelope  represented by a Gaussian  \eqref{gC}, and
$G_{mn}^H(\Omega)$ is a high-frequency function shown in Eq.~\eqref{gG}. A charge-noise spectrum $S_q(\omega)$ is given by Eqs.~\eqref{korE}.

The spectrum \eqref{spAx} has zero diagonal matrix elements: $S_{nn}(\omega)=0,$ since by definition $\tilde \Delta_{nn}=0$ and $q_{nn}=0.$ We also notice that
\ba \label{spAz}
S_{nm}(-\omega) = e^{-\omega/T}\, S_{mn}(\omega),
\ea
therefore,
\ba \label{spBx}
\tilde S_{nm}(-\omega) = e^{-\omega/T}\, e^{-\omega_{mn}/T}\,  \tilde S_{mn}(\omega).
\ea

\subsection{Evolution of system probabilities}

We see from Eq.~\eqref{probSig} that the probability $P_n(t)$ to observe the system in the state $\ket{n}$ is determined by the average value $\langle \sigma_n\rangle $ of the diagonal operator $\sigma_n = \sigma_{nn}.$
Averaging Eq.~\eqref{LanB} taken at $m{=}n$ over bath fluctuations leads to the equation
\ba \label{LanL1}
\langle \dot \sigma_n\rangle = i \sum_k \langle \tilde U^\dag \tilde Q_{nk} \ket{n}\bra{k} \tilde U\rangle + {\rm h.c.},
\ea
where ${\rm h.c.}$ means a Hermitian conjugate of the previous term. The unitary matrix $\tilde U$ is a functional of free bath variables $\tilde Q_{kl}$ as it is evident from Eq.~\eqref{Sch12}. Bath operators $\tilde Q_{kl}$ and $ Q_{kl}$ are characterized by non-Gaussian statistics as it follows from Eqs.~\eqref{Bath2} and \eqref{UnMat}. However, in the case of the weak system-bath coupling described by the non-diagonal Hamiltonian \eqref{Sch9} we can apply the quantum Furutsu-Novikov theorem \cite{ES1981,ASMA2018}
 to calculate average values in Eq.~\eqref{LanL1}. The goal is to remove a free-evolving bath operator from all terms in the equation for the system variables. To do that we have to pair this operator with other bath operators included into the evolution matrix $\tilde U:$
 \ba \label{LanL2}
 \langle \tilde U^\dag \tilde Q_{nk} \ket{n}\bra{k} \tilde U\rangle = \nn 
  \langle \wick{  \c {\tilde U}^\dag \c {\tilde Q}_{nk} \ket{n}\bra{k} \tilde U}\rangle +
  \langle \wick{   \tilde U^\dag \c {\tilde Q}_{nk} \ket{n}\bra{k} \c {\tilde U} }\rangle.
  \ea
  Here pairings between operators  are defined as 
 \ba \label{LanL3}
 \wick{ \c {\tilde U}^\dag(t) \c {\tilde Q}_{nk}(t) } {=} \int_0^\infty d t' \langle \tilde Q_{kn}(t') \tilde Q_{nk}(t)\rangle \frac{\delta \tilde U^\dag(t)}{\delta \tilde Q_{kn}(t')}, \;\nn
 \wick{ \c{\tilde Q}_{nk}(t)\ket{n}\bra{k} \c{\tilde U}(t) } = \nn \int_0^\infty dt' 
 \langle \tilde Q_{nk}(t) \tilde Q_{kn}(t')\rangle \ket{n}\bra{k}  \frac{\delta \tilde U(t)}{\delta Q_{kn}(t')}. \;
 \ea
A functional derivative $\delta /\delta Q(t')$ with respect to a bath variable $Q(t')$ actually means a derivative $\delta/\delta f(t')$ with respect to an auxiliary force $f(t'),$
 which is additive to $Q(t')$ in the system-bath Hamiltonian: 
\ba \label{funD1} 
  \frac{\delta}{\delta Q(t')} = \frac{\delta}{\delta f(t')}_{|f{=}0}. 
  \ea
  After taking the derivative the auxiliary force should be equated to zero: $f=0.$ 
 The main property of the functional derivative is that
 \ba \label{funD2}
 \frac{\delta Q_{mn}(t')}{\delta Q_{kl}(t')} = \delta_{mk}\, \delta_{nl} \,\delta(t-t'),
 \ea
 where $\delta_{mn}$ is the Kronecker delta, and $\delta(t-t')$ is the Dirac $\delta-$ function.
 
 Using these properties and also the definition \eqref{Sch12} of the matrix $\tilde U$ we calculate the functional derivative
 \begin{widetext}
 \ba \label{funD3}
 \frac{\delta \tilde U(t)}{\delta \tilde Q_{kl}(t')} =  i \mathcal T \Big\{ \int_0^t d \tau \delta(\tau {-} t') \ket{k(\tau)}\bra{l(\tau)}  \exp \Big[ i \sum_{m'n'} \int_0^t d \tau' \tilde Q_{m'n'}(\tau') \ket{m'(\tau')}\bra{n'(\tau')} \Big] \Big\} = \nn
 i \theta(t{-}t') \mathcal T \Big\{ \ket{k(t')}\bra{l(t')} \exp \Big[ i \sum_{m'n'} \int_0^t d \tau' \tilde Q_{m'n'}(\tau') \ket{m'(\tau')}\bra{n'(\tau')} \Big] \Big\} = \nn
 i \theta(t{-}t') \mathcal T \Big\{
 \exp \Big[ i \sum_{m'n'} \int_{t'}^t d \tau' \tilde Q_{m'n'}(\tau') \ket{m'(\tau')}\bra{n'(\tau')} \Big]\ket{k(t')}\bra{l(t')}\times \nn  \exp \Big[ i \sum_{m'n'} \int_0^{t'} d \tau' \tilde Q_{m'n'}(\tau') \ket{m'(\tau')}\bra{n'(\tau')} \Big] \Big\} =  i \tilde U(t,t') \ket{k(t')}\bra{l(t')} \tilde U(t') \, \theta(t-t'),
  \ea
 \end{widetext}
 where $\theta(t{-}t')$ is the Heaviside step function, and
 \ba \label{funD4}
 \tilde U(t,t') {=} \mathcal T \exp \Big[ i \sum_{mn} \int_{t'}^t d \tau \tilde Q_{mn}(\tau) \ket{m(\tau)}\bra{n(\tau)} \Big]\;
 \ea
 is the unitary matrix describing a time evolution between moments of time $t'$ and $t$. For clarity, we assume that basis states $\ket{k(t)}$ weakly depend on time $t$. 
 At $t>t'$ we have $$\tilde U(t,t') = \tilde U(t) \tilde U^\dag(t'),$$ therefore,
 \ba \label{funD5}
 \frac{\delta \tilde U(t)}{\delta \tilde Q_{kl}(t')}  = i \tilde U(t) \tilde U^\dag(t') \ket{k(t')} \bra{l(t')} \tilde U(t') = \nn i \tilde U(t) \,\sigma_{k l}(t') \, \theta(t-t').
 \ea
 In a similar way we obtain
 \ba \label{funD6}
 \frac{\delta \tilde U^\dag (t)}{\delta \tilde Q_{kl}(t')} = - i \sigma_{k l}(t') \,\tilde U^\dag (t)\, \theta(t-t').
 \ea
 Here we use the definition \eqref{sigMN} of the system operator 
 \ba \label{funD7}
 \sigma_{kl}(t') = \tilde U^\dag(t') \ket{k(t')}\bra{l(t')} \tilde U(t').
 \ea
 Taking into account Eqs.~\eqref{LanL1}, \eqref{LanL2}, \eqref{LanL3} and also Eqs.~\eqref{funD5}, \eqref{funD6} we find that a time derivative $\dot P_n {=} \langle \dot \sigma_n\rangle$ of the probability $P_n$ is determined by correlation functions of the system operators:
 \begin{widetext}
 \ba \label{LanL4} 
 \langle \dot \sigma_n\rangle = - \sum_m \int_0^t dt' \Big\{ \tilde K_{nm}(t,t') \langle \sigma_{nm}(t) \sigma_{mn}(t')\rangle + \tilde K_{nm}(t',t) \langle \sigma_{nm}(t') \sigma_{mn}(t)\rangle \Big\} \nn
 + \sum_m \int_0^t dt' \Big\{ \tilde K_{mn}(t,t') \langle \sigma_{mn}(t) \sigma_{nm}(t')\rangle + \tilde K_{mn}(t',t) \langle \sigma_{mn}(t') \sigma_{nm}(t)\rangle \Big\}.
 \ea
 \end{widetext}
 
 \subsection{Heisenberg-Langevin equations}
 
 In order to calculate correlation functions of system operators in Eq.~\eqref{LanL4} we have to rewrite Eq.~\eqref{LanB} in the form of a quantum Langevin equation \cite{ES1981}. This stochastic equation has a kinetic part, which determines relaxation in the system, and also a fluctuation force in the right-hand side. The bath average value of the fluctuation force should be equal to zero. A time derivative $\dot \sigma_{mn}$ in Eq.~\eqref{LanB} can be represented as
 \begin{widetext}
 \ba \label{LanM1}
 \dot \sigma_{mn} = \xi_{mn}  +
 i \sum_k \Big[ \wick{ \c{\tilde U}^\dag \c {\tilde Q}_{nk} \ket{m}\bra{k} \tilde U} -
 \wick{\c{\tilde U}^\dag \c{\tilde Q}_{km} \ket{k}\bra{n} \tilde U} \Big] + 
 i \sum_k \Big[ \wick{ \tilde U^\dag \c {\tilde Q}_{nk} \ket{m}\bra{k} \c{\tilde U}} -
 \wick{\tilde U^\dag \c{\tilde Q}_{km} \ket{k}\bra{n} \c{\tilde U}} \Big].
 \ea
 Pairings  between system and bath operators are shown in Eqs.~\eqref{LanL2}. A fluctuation force $\xi_{mn}$ has a straightforward definition
 \ba \label{LanXi}
 \xi_{mn} = i \sum_k \Big[ \tilde U^\dag \tilde Q_{nk} \ket{m}\bra{k} \tilde U  -
 \tilde U^\dag \tilde Q_{km} \ket{k}\bra{n} \tilde U \Big] \nn - 
 i \sum_k \Big[ \wick{ \c{\tilde U}^\dag \c {\tilde Q}_{nk} \ket{m}\bra{k} \tilde U} -
 \wick{\c{\tilde U}^\dag \c{\tilde Q}_{km} \ket{k}\bra{n} \tilde U} \Big]  - 
 i \sum_k \Big[ \wick{ \tilde U^\dag \c {\tilde Q}_{nk} \ket{m}\bra{k} \c{\tilde U}} -
 \wick{\tilde U^\dag \c{\tilde Q}_{km} \ket{k}\bra{n} \c{\tilde U}} \Big].
 \ea
 It follows from Eq.~\eqref{LanL2} that this force has zero mean value: $\langle \xi_{mn}\rangle =0.$ An explicit expression for $\xi_{mn}$ lets us to calculate correlation functions of fluctuation forces. 
 
 Taking into account definitions \eqref{LanL3} of system-bath pairings  we write Eq.~\eqref{LanM1} in the form 
 \ba \label{LanM2}
 \dot \sigma_{mn} = \xi_{mn} + i \int_0^\infty dt' \tilde K_{kn}(t',t) \frac{\delta \tilde U^\dag(t)}{\delta \tilde Q_{kn}(t')} \ket{m}\bra{k} \tilde U(t) - 
 i \int_0^\infty dt' \tilde K_{mk}(t',t) \frac{\delta \tilde U^\dag(t)}{\delta \tilde Q_{mk}(t')} \ket{k}\bra{n} \tilde U(t) \nn
 + i \int_0^\infty dt' \tilde K_{nk}(t,t')  U^\dag(t) \ket{m}\bra{k} \frac{\delta \tilde U(t)}{\delta \tilde Q_{kn}(t')} -
 i \int_0^\infty dt' \tilde K_{km}(t,t')  U^\dag(t) \ket{k}\bra{n} \frac{\delta \tilde U(t)}{\delta \tilde Q_{mk}(t')}.
 \ea
 Using here formulas \eqref{funD5} and \eqref{funD6} for functional derivatives of the unitary matrix $\tilde U$ and also a definition \eqref{sigMN}  we obtain a quantum Langevin equation for the system operator $\sigma_{mn},$
 \ba \label{LanM3}
 \dot \sigma_{mn} + \sum_k \int_0^t dt' \Big\{ \tilde K_{nk}(t,t')  \sigma_{mk}(t)  \sigma_{kn}(t') + \tilde K_{mk}(t',t)  \sigma_{mk}(t')  \sigma_{kn}(t) \Big\} \nn
 -\sum_k \int_0^t dt' \Big\{ \tilde K_{km}(t,t')  \sigma_{kn}(t)  \sigma_{mk}(t') + \tilde K_{kn}(t',t)  \sigma_{kn}(t')  \sigma_{mk}(t) \Big\} = \xi_{mn}.
 \ea
 \end{widetext}
 Averaging this equation (taken at $m{=}n$) over bath fluctuations produces Eq.~\eqref{LanL4} for the probability $P_n{=} \langle \sigma_n\rangle.$
 
 We are interested in a time evolution of the off-diagonal elements $\sigma_{mn}$ in the limit of weak system-bath coupling described by the non-diagonal Hamiltonian
 \eqref{Sch9}. In this case, in the Langevin equation \eqref{LanM3},   we can reduce system operators, such $ \sigma_{mk}(t),$ to those taken at the earlier moment of time $t'$ assuming that $\sigma_{mk}(t) \simeq \sigma_{mk}(t').$ 
 This assumption allow us to calculate products of system operators, such as 
 $$\sigma_{mk}(t) \sigma_{kn}(t') \simeq \sigma_{mk}(t') \sigma_{kn}(t') = \sigma_{mn}(t'), $$ and so on. The Langevin equation \eqref{LanM3} has now the simple form:
 \ba \label{LanM4}
 \dot \sigma_{mn} + \int_0^\infty \Sigma_{mn}(t,t') \,\sigma_{mn}(t') = \xi_{mn},
 \ea
 with the self-energy part
 \ba \label{LanM5}
 \Sigma_{mn}(t,t') = \sum_l [ \tilde K_{nl}(t,t') + \tilde K_{m l}(t',t)] \, \theta(t{-}t'),\;
 \ea
 and with the fluctuation force $\xi_{mn}$ defined by Eq.~\eqref{LanXi}.
 
 A self-energy part \eqref{LanM5} 
has a  Fourier transform: 
\ba \label{sigTx}
\Sigma_{mn}(\omega) = \int_{{-}\infty}^{{+}\infty} d\tau\; e^{-i\omega \tau}\, \Sigma_{mn}(\tau) = \nn \Sigma_{mn}'(\omega) {+} i\, \Sigma_{mn}''(\omega), 
\ea
 with real and imaginary components
\ba \label{sigUx}
\Sigma_{mn}'(\omega) = \sum_l \frac{ \tilde S_{nl}(\omega) + \tilde S_{ml}(-\omega)}{2}, \nn
\Sigma_{mn}''(\omega) = \sum_l \int \frac{d\Omega}{2\pi} \,  \left[ \frac{\tilde S_{nl}(\Omega)}{\omega - \Omega} + \frac{\tilde S_{ml}(\Omega)}{\omega + \Omega} \right].
\ea
The spectrum $\tilde S_{ml}(\omega)$ is defined by Eqs.~\eqref{LanK5} and \eqref{spAx}.
 We notice that
\ba \label{sigVx}
\Sigma_{nm}'(-\omega) = \Sigma_{mn}'(\omega), \; \Sigma_{nm}''(-\omega) = - \Sigma_{mn}''(\omega).
\ea
 
 In order to solve Eq.~\eqref{LanM5} we introduce the retarded Green function $\tilde {\mathcal G}_{mn}(t,t')$ satisfying the following equation:
 \ba \label{LanGreen}
 \frac{d}{d t} \tilde{\mathcal G}_{mn}(t{-}t') + \int_0^\infty dt_1 \,\Sigma_{mn}(t{-}t_1)\, \tilde{\mathcal G}_{mn}(t_1{-}t') \nn =   \delta(t-t').
 \ea
 
 The Green function $\tilde {\mathcal G}_{mn}(t,t_1)$ is defined by its Fourier image $\mathcal G_{mn}(\omega)$,
\ba \label{grA}
\mathcal G_{mn}(\omega) = \int d\tau e^{i\omega \tau}\, \tilde{\mathcal G}_{mn}(\tau) = \frac{1}{ - i\omega + \Sigma_{mn}(\omega)}, 
\ea
where $\tau = t - t_1.$ We note that
\ba \label{grB}
\mathcal G_{mn}(-\omega) = \mathcal G_{nm}^*(\omega). 
\ea
 With the Green function $\tilde{\mathcal G}_{mn}(t{-}t')$ the system operator $\sigma_{mn}(t) $ can be expressed in terms of the operator $\sigma_{mn}(t')$ taken at the earlier time $t'$ plus a contribution of the fluctuation force:
 \ba \label{LanSig1}
 \sigma_{mn}(t) = \tilde {\mathcal G}_{mn}(t{-}t') \, \sigma_{mn}(t') \nn + \int_{t'}^t dt_1 \, \tilde {\mathcal G}_{mn}(t{-}t_1)\, \xi_{mn}(t_1).
 \ea
 
 \vspace{0.25cm}
 
 \subsection{Master equations}

 The last term in the right-hand side of Eq.~\eqref{LanSig1} describes an effect of the fluctuation force during the correlation time $t-t'\sim \tau_c$ of the bath correlator $\tilde K_{mn}(t,t'). $ This effect is small and can be neglected. It means that
 \ba \label{LanSig2}
 \sigma_{mn}(t) \simeq \tilde{\mathcal G}_{mn}(t-t') \sigma_{mn}(t'), 
 \ea
 therefore,
 \ba \label{LanSig3}
  \sigma_{mn}(t) \sigma_{nm}(t') = \tilde{\mathcal G}_{mn}(t-t') \sigma_{m}(t'), 
  \ea
  and so on. 
  Substituting such correlators into Eq.~\eqref{LanL4} we derive a set of master equations for the probability distribution
  $\langle \sigma_n\rangle$ of the system over basis states $\ket{n}$:
  \begin{widetext}
  \ba \label{Mt1}
  \langle \dot \sigma_n\rangle +
  \sum_m \int_0^{t} dt' \Big[ \tilde K_{nm}(t,t') \,\tilde{\mathcal G}_{nm}(t-t') + K_{nm}(t',t) \,\tilde{\mathcal G}_{mn}(t-t')\Big] \langle \sigma_n(t')\rangle = \nn
   \sum_m \int_0^{t} dt' \Big[ \tilde K_{mn}(t,t') \,\tilde{\mathcal G}_{mn}(t-t') + K_{mn}(t',t)\, \tilde{\mathcal G}_{nm}(t-t')\Big] \langle \sigma_m(t')\rangle.
   \ea
   \end{widetext}
   In Eq.~\eqref{Mt1} moments of time $t$ and $t'$ are separated by a correlation time $\tau_c$ of the bath correlator $\tilde K_{mn}(t{-}t').$ A time evolution of the probability $P_n(t)=\langle \sigma_n(t)\rangle $ during this interval is insignificant. Therefore, in the collision terms of the master equation \eqref{Mt1} we can assume that $\langle \sigma_n(t')\rangle \simeq \langle \sigma_n(t)\rangle. $ With this assumption the master equation \eqref{Mt1} has the standard form
   \ba \label{Mt2}
   \dot P_n + \sum_n \Gamma_{mn} P_n = \sum_m \Gamma_{nm} P_m,
   \ea
   with the relaxation matrix
   \ba \label{Mt3}
   \Gamma_{mn} {=} \int_0^t d\tau [ \tilde K_{nm}(\tau) \tilde {\mathcal G}_{nm}(\tau) {+} \tilde K_{nm}({-}\tau) \tilde{\mathcal G}_{mn}(\tau)].\;
   \ea
  Considering that the running moment of time $t \gg \tau_c$ we assume in Eq.~\eqref{Mt3} that the upper limit $t=\infty.$ This allows us to write the rates $\Gamma_{mn}$ in the form
  \ba \label{Mt4}
  \Gamma_{mn} = \int \frac{d\omega}{2 \pi} \, \tilde S_{nm}(\omega)\, [ \mathcal G_{mn}(\omega) + \mathcal G^*_{mn}(\omega)], 
  \ea
  where $\tilde S_{nm}(\omega) = S_{nm}(\omega {-} \omega_{mn})$ is the shifted spectrum \eqref{spAx} of bath fluctuations, and $\mathcal G_{mn}(\omega)$ is the Fourier transform \eqref{grA} of the Green function $\tilde{\mathcal G}_{mn}(\tau). $
  The real and imaginary components of the self-energy function $\Sigma_{mn}(\omega)$ involved into Eq.~\eqref{grA} can be represented as combinations of linewidths $\gamma_n(\omega), \gamma_m(-\omega)$ and frequency shifts $\delta_n(\omega), \delta_m(-\omega)$ of $n$ and $m$ energy levels:
  \ba \label{gamDelta}
\Sigma_{mn}'(\omega) = \gamma_n(\omega) + \gamma_m(-\omega), \nn
\Sigma_{mn}''(\omega) = \delta_n(\omega) - \delta_m(-\omega),
\ea
where
\ba \label{gamXa}
\gamma_n(\omega) = \frac{1}{2} \sum_k S_{nk}(\omega+\omega_{nk}), \nn
\delta_n(\omega) = \sum_k \int \frac{d\Omega}{2\pi} \frac{S_{nk}(\Omega+\omega_{nk})}{\omega - \Omega}.
\ea
The frequency shift can be written also as
\ba \label{gamXb}
\delta_n(\omega) = \nn 
\sum_k \int_0^\infty \frac{d\Omega}{2 \pi} \frac{ S_{nk}(\omega {+} \omega_{nk}{-}\Omega) - 
S_{nk}(\omega {+} \omega_{nk}{+}\Omega) }{\Omega}.\;
\ea
With this information in mind we arrive at the final expression for the relaxation rates \eqref{Mt3}:
\ba \label{Mt5}
\Gamma_{mn} =  \int \frac{d\omega}{2\pi} \,S_{nm}(\omega - \omega_{mn})\times \nn   \frac{ 2 [ \gamma_n(\omega) + \gamma_m(-\omega)] }{ [ \omega - \delta_n(\omega)+\delta_m(-\omega)]^2 + 
  [ \gamma_n(\omega) + \gamma_m(-\omega)]^2 }.
  \ea
  \vspace{0.25 cm}
  
  In the limit of zero linewidths $\gamma_m, \gamma_n$ and zero frequency shifts $\delta_n, \delta_m$ the rate \eqref{Mt5} turns to the standard Bloch-Redfield form: $\Gamma_{mn} = S_{nm}(\omega_{nm}).$
The master equation \eqref{Mt2}
 describes both, equilibrium and nonequilibrium, situations.  
 In the equilibrium state we have
\ba \label{Mt6}
\dot P_n = \sum_m( \Gamma_{nm} P_m - \Gamma_{mn} P_n)= 0.
\ea
Taking into account a formula \eqref{Mt4} for $\Gamma_{mn},$ and also Eq.~\eqref{spBx} we find that
\ba \label{Mt7}
\Gamma_{nm} P_m - \Gamma_{mn} P_n = \nn 
2 \int \frac{d\omega}{2\pi} \tilde S_{mn}(\omega) {\mathcal G}'_{nm}(\omega) \Big[ P_m {-} e^{{-}\omega/T}  e^{{-}\omega_{mn}/T}\, P_n\Big],\;  
\ea
where ${\mathcal G}'_{nm}(\omega)$ is the real part of the Green function defined by Eq.~\eqref{grA},
\ba \label{grC}
{\mathcal G}'_{nm}(\omega) =  \nn \frac{   \gamma_m(\omega) + \gamma_n(-\omega) }{ [ \omega - \delta_m(\omega)+\delta_n(-\omega)]^2 + 
  [ \gamma_m(\omega) + \gamma_n(-\omega)]^2 }.
  \ea
  This function peaks at $\omega \simeq \delta_m(0)-\delta_n(0)$. Therefore, the equilibrium condition $\dot P_n{=}0$ given by Eq.~\eqref{Mt6} is satisfied provided that
  \ba \label{Mt8}
  \frac{P_n}{P_m} = \exp\Big( \frac{\tilde E_m - \tilde E_n}{T} \Big),
  \ea
  where $\tilde E_m$ and $\tilde E_n$ are shifted energy levels defined as $\tilde E_n = E_n + \delta_n(0)$. 
The ratio \eqref{Mt8} is specific for the Boltzmann distribution $P_n$ over shifted  energy levels  of the rf-SQUID. 
This means that the master equations \eqref{Mt2} is in agreement with the thermal equilibrium conditions.

\end{document}